\documentclass[a4paper]{article}

\usepackage{INTERSPEECH2020}
\usepackage{multirow}

\title{Unified Streaming and Non-streaming Two-pass End-to-end Model \\
for Speech Recognition}
\name{Binbin Zhang$^{1,3}$, Di Wu$^{1,3}$, Zhuoyuan Yao$^2$, Xiong Wang$2$, Fan Yu$2$, Chao Yang$^{1,3}$, Liyong Guo$^1$, Yaguang Hu$^1$, Lei Xie$^2$, Xin Lei$^1$}
\address{
  $^1$Mobvoi Inc., Beijing, China \\
  $^2$Audio, Speech and Language Processing Group (ASLP@NPU),
School of Computer Science, Northwestern Polytechnical University, Xi’an, China \\
  $^3$WeNet Open Source Community}

\email{binbinzhang@mobvoi.com}

\begin{document}

\maketitle
\begin{abstract}
In this paper, we present a novel two-pass approach to unify streaming and non-streaming end-to-end (E2E) speech recognition in a single model. 
Our model adopts the hybrid CTC/attention architecture, in which the conformer layers in the encoder are particularly modified. We propose a dynamic chunk-based attention strategy to allow arbitrary right context length. At inference time, the CTC decoder generates n-best hypotheses in a streaming manner. The inference latency could be easily controlled by only changing the chunk size. The CTC hypotheses are then rescored by the attention decoder to get the final result. This efficient rescoring process causes negligible sentence-level latency. Our experiments on the open 170-hour AISHELL-1 dataset show that, the proposed method can unify the streaming and non-streaming model simply and efficiently. On the AISHELL-1 test set, our unified model achieves 5.60\% relative character error rate (CER) reduction in non-streaming ASR compared to a standard non-streaming transformer. The same model achieves 5.33\% CER with 640ms latency in a streaming ASR setup.

\end{abstract}

\section{Introduction}

End-to-end (E2E) models have gained more and more attention to speech recognition over the last few years.
E2E models combine the acoustic, pronunciation and language models into a single neural network, showing competitive results compared to conventional ASR systems.
There are mainly three popular E2E approaches, namely CTC\cite{graves2006connectionist,amodei2016deep}, recurrent neural network transducer (RNN-T)\cite{graves2012sequence,graves2013speech} and attention based encoder-decoder (AED)\cite{chorowski2014end,chan2015listen,chorowski2015attention}.
They all have advantages and limitations in terms of recognition accuracy and application scenario, and many efforts have been paid for comparison of these models\cite{prabhavalkar2017comparison} or join some of them into one model\cite{kim2017joint,sainath2019two}.

% Streaming, RNN-T, MOcha, MMA, SCAMA，CTC-trigger
While these models have great performance in a none streaming application, it usually requires a lot of work or a lot of accuracy degradation to make the model work in a streaming way, and a lot of works have been done for that.
While RNN-T has the streaming ability in nature, with more suprior performance, AED models have to be modified to realize streaming function.
For RNN-T, a two pass\cite{sainath2019two,sainath2020streaming} method was proposed to close the accuracy gap to the non-streaming \textit{Listen, Attend, Spell}(LAS) model.
For AED, \textit{Hard Monotonic Attention}\cite{raffel2017online} is first proposed for monotonically align the input and output of the AED model, and then it could work in a streaming way. With the same idea, \textit{Monotonic Chunkwise Attention}(MoChA)\cite{chiu2017monotonic} and \textit{Monotonic Multihead Attention}\cite{inaguma2020enhancing} were proposed to further improve performance and stability of monotonic attention. 

% Unified model and % Our work
Recently there have been also increasing interests in unifying non-streaming and streaming speech recognition models into one model. Some transducer based models such as Y-model \cite{tripathi2020transformer} and UNIVERSAL ASR\cite{yu2020universal} have been designed for this goal with good performance. The unified model not only reduces the accuracy gap between the streaming model and the non-streaming counterpart, but also alleviates the burden of model development, training and deployment. 

In this work, we propose a new framework namely U2 to unify non-streaming and streaming speech recognition. Our framework is based on the hybrid CTC/attention architecture with conformer blocks. The training process is simple and it avoids the RNN-T model's complicated tricks and instability issues. To support streaming, we modify the conformer block while bringing negligible performance degradation. In further, by using a dynamic chunk training strategy, our framework allows users to control the latency at inference time. Our results show that U2 achieves state-of-the-art streaming accuracy on the public Aishell-1 dataset\cite{bu2017aishell}.

\section{Related Works}

% Joint CTC training
Hybrid CTC/attention end-to-end ASR in \cite{kim2017joint} adopted both CTC and attention decoder loss in training to achieve fast convergence and to improve robustness of the AED model. However, during decoding, it combined attention score and CTC score and performs joint decoding. Both of the scores can only be computed until the whole speech utterance is available, which makes it apparently a non-streaming model.

% RNN-T two pass
Two pass based model\cite{sainath2019two}  was proposed on RNN-T and achieves comparable accuracy to a LAS model. 
However, RNN-T training is very memory consuming\cite{li2019improving} so that we can not use a large batch for training on typical GPUs, which results in a very slow training speed as well as poor performance. Besides, RNN-T training is also unstable. CTC pre-training  \cite{rao2017exploring} and \textit{Cross Entropy}(CE) pre-training were proposed in \cite{hu2020exploring} to assist RNN-T training, while pre-training was also tricky and complicated. These increase the difficulty of using RNN-T in speech recognition application, especially when lacking computing and research resources.
And it was pointed out that training directly from scratch is unstable using combined RNN-T loss and LAS loss, so a three-step training strategy was proposed in \cite{sainath2019two} to solve the problem, which further complicates the training pipeline.

% Model-Y and Universal ASR
For unified non-streaming and streaming model,
Y-model uses variable context at training and several optional contexts at inference. However, the optional contexts are predefined at the training stage, and the contexts have to be carefully designed in terms of the number of encoder layers, the kernel size of the convolution operation.
Moreover dual-mode only has one streaming configuration in both training and inference. If we want another streaming model with different latency at inference, the model needs to be totally retrained. Besides, both Y-model and dual-mode are RNN-T based models. They have the same drawbacks as RNN-T.

Our proposed U2, a CTC-AED joint model, is trained by combined CTC and AED loss and dynamic chunk attention. It not only unifies the non-streaming and streaming model, giving promising result, but also significantly simplifies the training pipeline, as well as dynamically controls the trade-off between latency and accuracy in streaming applications.

\section{U3}

\subsection{Model architecture}

\begin{figure}[h]
  \centering
  \includegraphics[width=\linewidth]{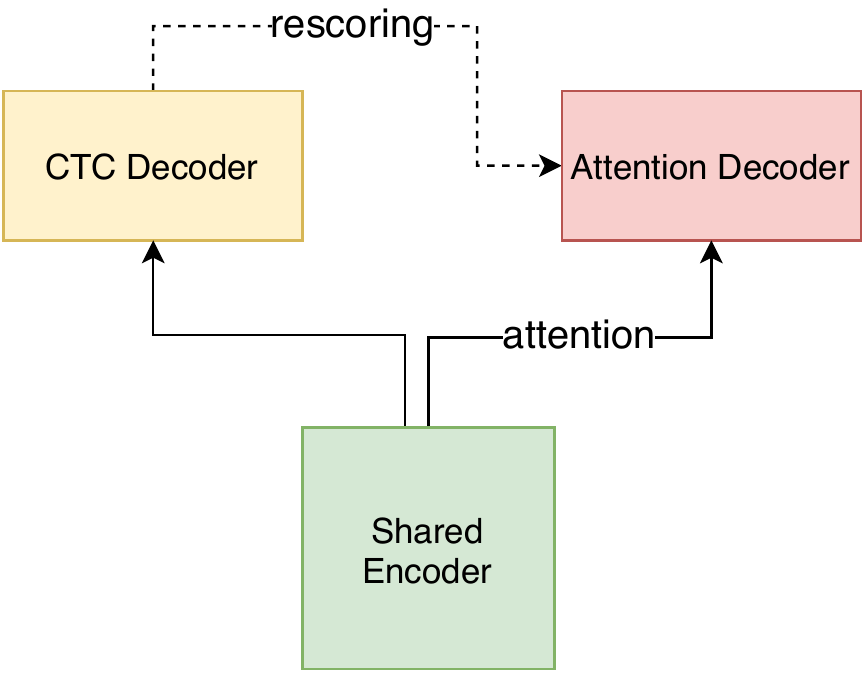}
  \caption{Two pass CTC and AED joint architecture}
  \label{fig:ctc_attention_joint}
  \vspace{-1.5em}
\end{figure}

The proposed three-pass architecture is shown in Figure \ref{fig:ctc_attention_joint}. It contains three parts, a \textit{Shared Encoder}, a \textit{CTC Decoder} and a \textit{Attention Decoder}. The \textit{Shared Encoder} consists of multiple Transformer\cite{vaswani2017attention} or Conformer\cite{gulati2020conformer} encoder layers. The \textit{CTC Decoder} consists of a linear layer and a log softmax layer, The CTC loss function is applied over the softmax output in training. The \textit{Attention Decoder} consists of multiple Transformer decoder layers. We can make the \textit{Shared Encoder} only see limited right contexts, then CTC decoder could run in a streaming mode in the first pass. In the second pass, the output of the \textit{Shared Encoder} and \textit{CTC Decoder} can be used in different ways. The training and decoding processes are detailed in the following.
\subsection{Training}
\subsubsection{Combined Loss}

The training loss is combined with CTC loss and AED loss as listed in  \ref{eq:combined_loss}, where $\mathbf{x}$ is the acoustic feature, $\mathbf{y}$ is the corresponding annotation, $\mathbf{L}_{\text{CTC}}\left(\mathbf{x}, \mathbf{y}\right)$,
$\mathbf{L}_{\text{AED}}\left(\mathbf{x}, \mathbf{y}\right)$ are the CTC and AED loss respectively,
$\lambda$ is a hyperparameter which balance the importance of CTC and AED loss. Unlike RNN-T based two pass in \cite{sainath2019two} where a three step process was used to stable the training, we can directly train our model by the combined loss from scratch, which significantly simplify our training pipeline. And As shown in \cite{kim2017joint}, the combined loss also help the model converge faster and have better performance.
\begin{equation}
  \label{eq:combined_loss}
    \mathbf{L}_{\text {combined}}\left(\mathbf{x}, \mathbf{y}\right)=\lambda \mathbf{L}_{\text{CTC}}\left(\mathbf{x}, \mathbf{y}\right)+(1-\lambda) (\mathbf{L}_{\text{AED-L }}\left(\mathbf{x}, \mathbf{y}\right) + \mathbf{L}_{\text{AED-R }}\left(\mathbf{x}, \mathbf{y}\right))
\end{equation}

\subsubsection{Dynamic Chunk Training}

A Dynamic chunk training technique is proposed in this section to unify the none streaming and streaming model and enable latency control. 

\begin{figure}[h]
  \centering
  \includegraphics[width=\linewidth]{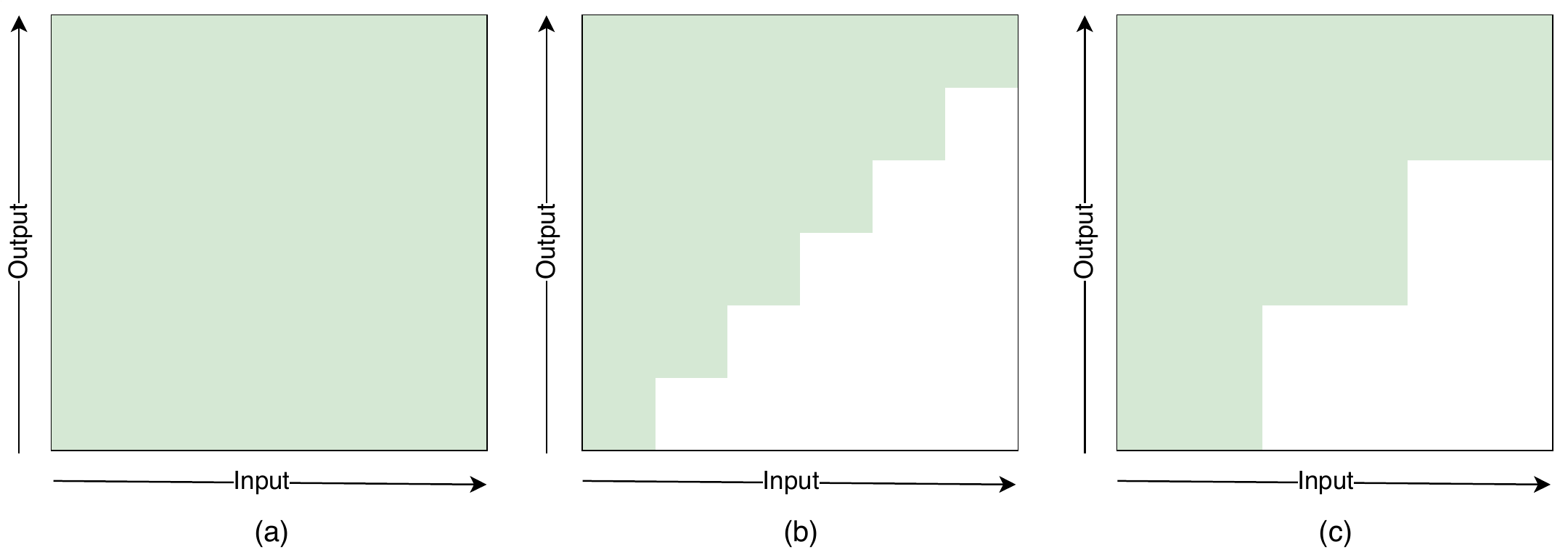}
  \caption{Full attention, Left attention, Chunk Attention}
  \label{fig:chunk_attention}
  \vspace{-1.0em}
\end{figure}

As described before, our U2 could only be streaming when the \textit{Shared Encoder} is streaming. Full self attention is used in standard Transformer encoder layers, as shown in Figure \ref{fig:chunk_attention} (a), every input at time $t$ depends on the whole inputs, green means there is a dependency, while white means there is no dependency.
The simplest way to stream it is to make the input $t$ only see itself and the input before $t$, namely left attention, seeing no right context, as shown in Figure \ref{fig:chunk_attention} (b), but there is very big degradation compared to full context model.
Another common technique is to limited input $t$ only see a limited right context {$t+1,t+2,...,t+W$}, where $W$ is the right context for each encoder layer, and the total context is accumulated through all the encoder layers, for example, if we have $N$ encoder layers, each has $W$ right context, the total context is $N*W$. Right context usually improves performance compared to pure left attention, however, we should carefully design the number of layers and every right context for each layer to control the final right context for the whole model, and things get more difficult when we use conformer encoder layer, in which convolution through time with right context is used.
We adopt a chunk attention in this work, as shown in Figure \ref{fig:chunk_attention} (c), we split the input to several chunks by a fixed chunk size $C$, the dark green is for the current chunk, for each chunk we have inputs [t+1, t+2, ..., t+C], every chunk depends on itself and the all the previous chunks. Then the whole latency of the encoder depends on the chunk size, which is easy to control and implement. We can train the model using a fixed chunk size, we call it static chunk training, and decoding with the same chunk.

Motivated by the idea of unified E2E model, we further propose a dynamic chunk training. We can use dynamic chunk size for different batches in training, the dynamic chunk size range is a uniform distribution from 1 to max utterance length, namely the attention varies from left context attention to full context attention, and the model captures different information on various chunk size, and learns how to do accurate prediction when different limited right context provided. We call the chunks which sizes are from 1 to 25 as streaming chunk for streaming model and size which is max utterance length as none streaming chunk for none streaming model. However, the results of this method is not good enough, so next we change the distribution of chunk size during training process as follows.
\begin{equation}
\label{eq:half distribution}
chunk size =
\begin{cases}
l_{max} & x > 0.5\\
l \sim  U(1, min(25, l_{max}-1)) & x \leq 0.5
\end{cases}
\end{equation}

As shown in Equation 2, x is sampled from 0 to 1.0 in each batch during the training process, l$_{max}$ is the max utterance length of current batch, and U is a uniform distribution. So the distribution of chunk size changed, half is full chunk for none streaming, and the other half from 1 to 25 is used for streaming. 

Our later experiments will show, this is a simple but efficient way, the model trained by dynamic chunk size has a comparable performance compared to static chunk training.

Besides this batch level method, we also tried epoch level - using full chunk for the first half epochs and streaming chunk for the second half or in turn. But these strategies do not work.

\subsubsection{Causal Convolution}

The convolution units in conformer consider both left and right context. The total right context depends on  convolution layer's context and the stack number of conformer layers. So this structure not only brings in  additional latency, but also ruin the benefits of chunk-based attention, that the latency is independent on the network structure and could be just controlled by chunk at inference time. To overcome this issue, we use casual convolution\cite{tripathi2020transformer} instead.

\subsection{Decoding}

The \textit{Shared Encoder} consumes the audio feature chunk by chunk. The larger chunk size usually means higher latency and better accuracy and the maximum latency is proportional to the frame number of one chunk. The proper decoding chunk size depends on specific task requirements.

The \textit{CTC Decoder}  outputs first pass hypotheses in a streaming way. At the end of the input, the \textit{Attention Decoder} uses full context attention to get better results. Two different modes are explored here:
\begin{itemize}
\item Attention Decoder mode. The CTC results are ignored in this mode. \textit{Attention Decoder} generate outputs in an auto-regressive way with the attention of the output of \textit{Shared Encoder}. 

\item Rescoring mode. The n-best hypotheses from CTC are scored by the \textit{Attention Decoder} with the output of the \textit{Shared Encoder} in a teacher-forcing mode. The best re-scored hypothesis is used as the final result. This mode avoids the auto-regressive process and achieves better real-time factor(RTF). Besides, the CTC scores could be weighted combined to get a better result in a simple way. 
\end{itemize}

\begin{equation}
  \label{eq:combined_scores}
    SCORES_{\text {final}} =\lambda*SCORES_{\text {CTC}} + SCORES_{\text {attention}}
\end{equation}

In order to get a better result, ctc weighted score was added during rescoring mode decoding as shown in Equation 3, and our later experiments will show that it is always beneficial to decoding results. 

\section{Experiments}

In order to evaluate our proposed U2, we carried out our experiments on the open-source Chinese Mandarin speech corpus AISHELL-1\cite{bu2017aishell}, which contains a 150-hour training set, a 10-hour develoment set and a 5-hour test set. The test set contains 7176 utterances in total.
We use wenet\footnote{https://github.com/wenet-e2e/wenet} end-to-end speech recognition toolkit for all experiments.

We use the state-of-the-art ASR network---Conformer \cite{gulati2020conformer} as our shared encoder, and the decoder part is the same as the traditional transformer decoder. Conformer adds convolution module on the basis of transformer so that it can model both local and global context and results in better results on different ASR tasks. As for the dynamic chunk training of the conformer model, causal convolution is used instead in the experiments making our encoder is independent to the right context. \\

\subsection{AISHELL-1 Task}
For AISHELL-1, we use 80 dimensional log-mel filter bank (FBank) splice 3 dimensional pitch computed on 25ms window with 10ms shift as feature. And we do speed perturb with 0.9, 1.0 1.1 on the whole data to generate 3-fold speed changes. 
SpecAugment\cite{park2019specaugment} is applied with 2 frequency masks with maximum frequency mask ($F = 10$), and 2 time masks with maximum time mask($T=50$). Two convolution sub-sampling layers with kernel size 3*3 and stride 2 is used in the front of the encoder, namely 4 times sub-sampling in total. For encoder, we use 12 conformer layers with 4 multi head attention. For the \textit{Attention Decoder}, we use 6 transformer layers with 4 multi head attention. Each conformer layer uses 256 attention dimension and 2048 feed forward dimension. Accumulating grad was also used to stabilize training, and we update parameters every four steps. Attention dropout, feed forward dropout and label smoothing regularization are applied in each encoder and decoder layer in order to prevent over-fitting. We use Adam optimizer and transformer learning rate schedule with 25000 warm-up steps to train models. Moreover, we get our final model by averaging the top 10 best models which have a lower loss on the dev set at the training.

\begin{table}[h]
    \centering
    \setlength{\abovecaptionskip}{0.1cm}
    \caption{Decoding method comparison}
    \label{tab:decoding_method}
    \begin{tabular}{lllll}
        \toprule
        \textbf{decoding method}        &
        \textbf{CTC weight}             &
        \textbf{RTF}                    &
        \textbf{CER}           \\ \midrule
        attention decoder      & / & 0.197  & 4.92      \\
        ctc prefix beam search & / & /       & 4.93      \\
        attention rescoring   & 0.0 & /     & 4.72 \\
        attention rescoring   & 0.5 & \textbf{0.082} & 4.64  \\
        \bottomrule
    \end{tabular}
    \vspace{-1.5em}
\end{table}
% alPlease add the following required packages to your document preamble:
% \usepackage{multirow}
% \usepackage{graphicx}
\newcommand{\tabincell}[2]{\begin{tabular}{@{}#1@{}}#2\end{tabular}}  
\begin{table*}[h]
    \setlength{\abovecaptionskip}{0.1cm}
    \caption{Dynamic vs static chunk training}
    \centering
    \label{tab:dynamic_vs_static}
    \begin{tabular}{c l l l l l l}
    \hline
    \multirow{2}{*}{training method}                                & \multirow{2}{*}{decoding mode} & \multicolumn{5}{c}{decoding chunk size} \\ \cline{3-7} 
                                                                    &                                & full   & 16     & 8     & 4     & 1      \\ \hline
    \multirow{3}{*}{static chunk training, static chunk inference}  & attention decoder              & 5.35   & 5.95   & 5.99  & 6.15  & 6.36   \\ \cline{2-7} 
                                                                    & ctc prefix beam search         & 5.18   & 6.30   & 6.50  & 6.69  & 6.73   \\ \cline{2-7} 
                                                                    & attention rescoring            & \textbf{4.86}   & 5.55   & 5.78  & 6.06  & 6.02    \\ \hline
    \multirow{4}{*}{\shortstack{ dynamic chunk training, static chunk inference }} & attention decoder              & 5.27   & 5.51   & 5.67  & 5.72  & \textbf{5.88}   \\ \cline{2-7} 
                                                                    & ctc prefix beam search         & 5.49   & 6.08  & 6.41  & 6.64  & 7.58  \\ \cline{2-7} 
                                                                    & attention rescoring            & 4.90   & \textbf{5.33}  & \textbf{5.52}  & \textbf{5.71}  & 6.23   \\ \hline
    \end{tabular}%
    \vspace{-1.5em}
\end{table*}
% Decoding method comparison, and show error pattern.
\subsubsection{Decoding Method}

First, we explore different decoding methods on a none streaming model, in which full context and a conformer with standard convolution kernel size 15 are used in training, to ensure both CTC and AED decoder give a reasonable result. For AED decoder, we use beam 10 for decoding. We use prefix beam search for CTC, which is used for generating top-n different hypothesises for later rescoring.

As shown in the Table \ref{tab:decoding_method}, the attention rescoring result outperforms both CTC prefix beam search and attention decoder results, which is out of our expectation.

After analyzing the decoding results of CTC prefix beam search and attention rescoring, we found that a lot of wrong results generated by CTC prefix beams search could be corrected by attention rescoring, However some good cases in were false corrected after attention rescoring, which means CTC plays an important role in some cases. So we added CTC weight during attention rescoring as Equation \ref{eq:combined_scores}. And we tested different CTC weights from 0.1 to 0.9, all of them helps attention rescoring in our experiments, and 0.5 is the most stable one. Here we just show the result when CTC weight is 0.5, as we see, when combining with CTC weight, the CER can be further reduced to 4.72.
To our knowledge, it's the best published result on AISHELL-1. And 0.5 is the default CTC weight of attention rescoring mode in our later experiments. 

Since standard attention decoder is running in an autoregressive fashion, which is time consuming, while attention rescoring just uses attention decoder for rescoring, it can be processed in parallel, and it should be faster in theory.
So here we also investigate the RTF of both attention decoder and attention rescoring method, and single thread is used during decoding in Pytorch. As expected in Table \ref{tab:decoding_method}, we got 2.40 times speed up by attention rescoring compared to attention decoder in decoding.

To conclude here, we can see the attention rescoring is both faster and more accurate.

\subsubsection{Dynamic chunk evaluation}
% Dynamic chunk evaluation
As mentioned before, causal convolution is used in dynamic chunk training to unify none streaming and streaming model, and a kernel size of 8 is used, which is half of the previous experiment since the model is limited to see left context only here.

In order to compare with static chunk training, we trained the five different models with different static chunk size full/16/8/4/1, and then decode with the same chunk size as our baseline. And we trained only one unified model with the aforementioned dynamic chunk strategy as in Equation \ref{eq:half distribution}. The result is shown in Table \ref{tab:dynamic_vs_static}, and we mainly pay our attention to the attention rescoring result here since it's the final performance of our system. As we can see from the table, dynamic chunk trained model has a little degradation on full chunk and chunk 1, which are the two boundary points of the dynamic chunk with infinite latency and no latency respectively. We guess it's more difficult to learn boundary information in the unified model.
However, we see a slight gain over static chunk trained model when chunk size is 16/8/4, which means dynamic chunk strategy benefits the unified model by varying chunk training in this case. 
\begin{table}[h]
    \setlength{\abovecaptionskip}{0.1cm}
    \centering
    \caption{Comparison to other streaming solutions}
    \label{tab:solution_comparison}
    \begin{tabular}{llll}
    \toprule
    model            & params(M) & latency(ms) & CER  \\ \midrule
    Sync-Transformer\cite{tian2020synchronous} & /      & 400   & 8.91 \\
    SCAMA\cite{zhang2020streaming}            & 43    & 600   & 7.39 \\
    MMA\cite{inaguma2020enhancing}              & /      & 640   & 6.60  \\
    U2          & 47    & \textbf{320}+$\Delta$  & \textbf{5.33} \\ \bottomrule
    \end{tabular}
    \vspace{-1.5em}
\end{table}

\begin{table}[t]
    \centering
    \setlength{\abovecaptionskip}{0.1cm}
    \caption{Comparison U2 and static full attention on a 15000-hour Mandarin speech recognition task}
    \label{tab:15000 hour}
    \begin{tabular}{c c c c }
    \hline
    \multirow{2}{*}{test set}  &
    \multirow{2}{*}{EXP1} & \multicolumn{2}{c}{EXP2} \\ \cline{3-4} 
                                                                    &                                & full   & 16          \\ \hline
    \multirow{1}{*}{aishell}  
                                                                    & 3.96            & \textbf{3.70}   & 4.41    \\ \hline
    \multirow{1}{*}{tv}     & \textbf{10.92}            & 11.96   & 13.51    \\ \hline
    \multirow{1}{*}{conversation}     & \textbf{12.95}            & 14.01   & 15.35    \\ \hline
    \end{tabular}%
    \vspace{-1.5em}
\end{table}

Overall, the dynamic chunk trained model is comparable static chunk trained models, so we can easily unify the none streaming model and streaming model into one single model by our U2 framework via the two pass decoding and dynamic chunk training.

\subsubsection{Comparison to other solutions}
  
Table \ref{tab:solution_comparison} lists several published streaming solutions on AISHELL-1 test set, including Sync-Transformer\cite{tian2020synchronous}, SCAMA\cite{zhang2020streaming},  and MMA\cite{inaguma2020enhancing}. $\Delta$ is the additional latency introduced by attention rescoring at the end of decoding in our U2, but it's fast enough as we have talked before it can be paralleled into one batch computing, and it's 50-100ms as analysed in \cite{sainath2019two,tripathi2020transformer}. We can see our U2 has far surpassed other solutions with a small additional latency.\\

\subsection{15,000-hour Tasks}

We extend our experiments on a mixed 15,000-hour dataset which collected several domains, all in Mandarin, including variety show, talk show, tv soap, podcast, and radio.
The same acoustic features as mentioned in Section 4.1 was used.
First, we trained a conventional full attention conformer model which uses the same layers mentioned in 4.1 but uses 384 attention units.
For the second experiment, we trained a u2 model which parameters are the same as the first experiment using the method mentioned in Section 3. 
Three test set was used to evaluate models including AISHELL-1, tv domain, and conversation domain. The results are reported in the \label{tab:15000 hour}.
U2 gets comparable results to EXP1 baseline and even better on AISHELL-1 test set when using full attention during inference. 
Though both of convolution and self-attention in conformer encoder was limited to the current and left context when chunk size is 16 during inference, CER does not appear obvious decay.

\section{Conclusions}
 We propose a framework to train a single model which can do recognition in both streaming and full context way. This framework can be trained directly and stably without complicated training process. A fast weighted re-score method is used to get full-context performance with little additional latency. We also propose a dynamic chunk based strategy to improve the model performance and enable trading off the latency and accuracy conveniently at inference time. 

\newpage

\bibliographystyle{IEEEtran}
\bibliography{main}

\end{document}